\documentclass[aps,pre,reprint,superscriptaddress,showpacs,preprintnumbers,amsmath,amssymb]{revtex4-1}
\usepackage{graphicx}
\graphicspath{{fig/}}
\usepackage{dcolumn}
\usepackage{bm}
\usepackage{color}
\begin{document}
\preprint{Preprint}
\title{The role of microscopic friction in statistics and scaling laws of avalanches}
\author{Kuniyasu Saitoh}
\affiliation{Department of Physics, Faculty of Science, Kyoto Sangyo University, Motoyama, Kamigamo, Kita-ku, Kyoto 603-8555, Japan}
\date{\today}
\begin{abstract}
We investigate statistics and scaling laws of avalanches in two-dimensional frictional particles by numerical simulations.
We find that the critical exponent for avalanche size distributions is governed by microscopic friction between the particles in contact,
where the exponent is larger and closer to mean-field predictions if the friction coefficient is finite.
We reveal that microscopic ``slips" between frictional particles induce numerous small avalanches
which increase the slope, as well as the power-law exponent, of avalanche size distributions.
We also analyze statistics and scaling laws of the avalanche duration and maximum stress drop rates, and examine power spectra of stress drop rates.
Our numerical results suggest that the microscopic friction is a key ingredient of mean-field descriptions
and plays a crucial role in avalanches observed in real materials.
\end{abstract}
%
\maketitle
%
\section{Introduction}
\label{sec:intro}
Amorphous solids,\ e.g.\ granular materials and glasses, are ubiquitous in nature
and understanding of their mechanical properties is crucial to engineering science \cite{lemaitre}.
When amorphous solids are continuously sheared, one observes that the system reaches a steady state after the yielding \cite{yield_t0,yield_t1,yield_t2,yield_t3},
where the increase of stress is (suddenly) truncated by a plastic event such that the stress fluctuates around its mean value.
Such a plastic or stress drop event is often called \emph{avalanche} in the literature \cite{dahmen1,aval_review0}
and its statistical properties have been widely investigated in the context of non-equilibrium phase transition \cite{noneq_phase} such as the self-organized criticality \cite{review4}.

Among many theoretical studies, conventional mean field (MF) descriptions have well explained statistics of avalanches in amorphous solids \cite{aval_mean3,aval_mean2,aval_mean0,aval_mean1}.
These MF approaches are based on the analogy between avalanche and the \emph{depinning transition} of elastic faults in heterogeneous media,
where interactions between dislocation pairs are considered as long-ranged \cite{aval_mean3,aval_mean2,aval_mean0,aval_mean1}.
Strikingly, power-law distributions of the size of avalanche,\ i.e.\ $P(S)\sim S^{-\tau}$, are predicted by the MF theories,
where the MF prediction, $\tau=3/2$, is confirmed in a broad range of materials, from nanocrystals to earthquakes \cite{aval_mean8,dahmen0,dahmen1,dahmen2,aval_mean11}.
The MF theories also predict power-law distributions of avalanche duration (i.e.\ life time of a stress drop event) as $P(T)\sim T^{-\kappa}$,
where the theoretical prediction $\kappa=2$ is observed in experiments \cite{aval_mean6}.
In addition, the scaling of power spectra of stress drop rates is described as $P(\omega)\sim\omega^{-2}$ in the MF theories,
which is well reproduced by the experiments of bulk metallic glasses \cite{aval_mean7} and astrophysical objects \cite{aval_mean6}.
Furthermore, scaling laws of the avalanche size $S$, avalanche duration $T$, and maximum stress drop rate $M$ are given by $T\sim S^{1/2}$ and $M\sim S^{1/2}$ in the MF descriptions,
where both well agree with experimental results \cite{aval_mean6,aval_mean7,aval_mean5}.

On the other hand, the power-law exponent $\tau$ evaluated by numerical simulations tends to be smaller than the MF prediction ($\tau=3/2$).
For instance, an extremely small value for distributions of potential energy drop,\ i.e.\ $P(\Delta U)\sim \Delta U^{-\tau}$ with $\tau=0.7$,
was found in molecular dynamics (MD) simulations of two-dimensional foam under shear \cite{aval_foam0,aval_foam1}.
In addition, $\tau=1$ for avalanche size distributions was reported by athermal quasi-static (AQS) simulations of two-dimensional soft particles,
where interactions between the particles in contact are modeled by a harmonic potential \cite{aval_quasi1}.
Moreover, the exponent seems to be sensitive to particle properties:
Two exponents, $\tau=1.15$ and $1.25$, for $P(S)$ are obtained from different types of the Lennard-Jones potentials \cite{aval_finite0},
while it can be controlled in the range $1\leq\tau\leq 1.5$ by the effect of particle inertia \cite{aval_inert0,aval_inert1}.
These numerical results indicate that the concept of \emph{universality class} of avalanches \cite{dahmen1} is violated
in the sense that the critical exponent $\tau$ depends on microscopic details.

In addition to the atomistic (MD and AQS) simulations,
mesoscopic elastoplastic (EP) models \cite{aval_review0} predict the exponent in the range $1.2\leq\tau\leq 1.35$
depending on the spatial dimensions \cite{aval_inert2,aval_other0,aval_rate0,aval_scaling0,aval_scaling1,aval_epmodel0,aval_epmodel1}.
Different from the conventional MF approaches, the EP models employ an anisotropic propagator in the governing equation of local stress,
where the propagator is long-ranged and obeys quadrupolar symmetry in the shear plane (as Eshelby's inclusion) \cite{aval_review0}.
Then, on the basis of the analogy between avalanche and the \emph{yielding transition}, Lin et al. suggested \emph{scaling relations},
where the exponent $\tau$ is given by the spatial dimension $d$, fractal dimension $d_f$,
and power-law exponent for distributions of local distance to yield stress $\theta$ \cite{aval_scaling0,aval_scaling1}.
Note that the recent study of discrete dislocation dynamics also pointed out dissimilarities
between the MF descriptions (i.e.\ the analogy to the depinning) and avalanches \cite{aval_dislocation}.
Furthermore, recent experiments and simulations of two-dimensional granular materials reported
similar power-law distributions of avalanche energy,\ i.e.\ $P(E)\sim E^{-1.24}$ and $E^{-1.43}$, respectively \cite{aval_frictional}.

Because of the discrepancy in the MF theories, experiments, atomistic simulations, and mesoscopic models,
many factors,\ e.g.\ system sizes \cite{aval_finite0,aval_finite1,aval_finite2}, driving rates \cite{aval_rate0},
and temperature \cite{aval_finite3,aval_finite4}, have been examined by numerical studies of avalanches.
However, much less attention has been paid to the effect of \emph{microscopic friction},\
i.e.\ the dynamical or Coulomb friction between the particles in contact \cite{dem}.
The microscopic friction is intrinsic to amorphous solids in nature,\ e.g.\ granular materials,
and thus it is important to clarify how the microscopic friction affects the statistics of avalanches (especially the exponent $\tau$) and scaling laws.

In this paper, we numerically investigate statistics and scaling laws of avalanches in two-dimensional frictional particles.
By using the \emph{frictional contact model} \cite{dem}, we implement the microscopic friction into MD simulations.
We find that the statistics and scaling laws approach the MF predictions if the friction coefficient is finite.
Interestingly, the exponent $\tau$ is well described by the scaling relations proposed by Lin et al. \cite{aval_scaling0,aval_scaling1}
if the microscopic friction is absent,\ i.e.\ if the system is \emph{frictionless}.
However, they do not hold in frictional systems.
We show that the drop of tangential stress contributes to the population of small avalanches, which eventually increases the exponent $\tau$ in frictional systems.

In the following, we explain our numerical methods in Sec.\ \ref{sec:method} and show our numerical results in Sec.\ \ref{sec:result}.
In Sec.\ \ref{sec:discuss}, we discuss and conclude our findings.
%
\section{Methods}
\label{sec:method}
We study two-dimensional soft athermal particles by MD simulations.
To avoid crystallization, we use $50:50$ binary mixtures of $N$ particles,
where different kinds of particles have the same mass $m$ and different diameters (their ratio is $1.4$) \cite{saitoh14}.
The force between the particles in contact is divided into normal and tangential directions \cite{dem}:
The normal force $f_n$ is modeled by a linear spring-dashpot, where the spring constant and viscosity coefficient are given by $k_n$ and $\eta_n$, respectively.
The tangential force $f_t$ is also described by the linear spring-dashpot with the same spring constant and viscosity coefficient,\ i.e.\ $k_t=k_n$ and $\eta_t=\eta_n$, respectively.
However, it switches to dynamical friction $\mu|f_n|$ with the friction coefficient $\mu$ if it exceeds a threshold,\ i.e.\ if $|f_t|>\mu|f_n|$.
We adjust the spring constant and viscosity coefficient as the normal restitution coefficient is given by $e=\exp(-\pi/\sqrt{2k_n\eta_n^{-2}-1})\simeq0.7$ \cite{dem}.
In addition, we introduce a time unit as $t_0\equiv\eta_n/k_n=\eta_t/k_t$.

We randomly distribute the $N$ particles in an $L\times L$ square periodic box, where the area fraction of the particles is given by $\phi=0.9$.
We then apply simple shear deformations to the system under the Lees-Edwards boundary condition \cite{lees}.
In each time step, every particle position $\bm{r}_i=(x_i,y_i)$ is replaced with $(x_i+\delta\gamma y_i,y_i)$ ($i=1,\dots,N$)
and equations of translational and rotational motions are numerically integrated with a small time increment $\delta t$ \cite{daniel}.
Here, we use $\delta\gamma=10^{-7}$ for the strain increment.

In the following analyses, we control the parameters, $\mu$ and $N$, and scale every mass, time, and length by $m$, $t_0$, and the mean particle diameter $d_0$, respectively.
In addition, we only analyze the data in a steady state, where the amount of shear strain exceeds unity $\gamma>1$.
%
\section{Results}
\label{sec:result}
In this section, we show our numerical results of stress-strain curves (Sec.\ \ref{sub:stress}) and explain stress drop events (Sec.\ \ref{sub:drop}).
We examine the MF predictions of statistics of avalanches (Sec.\ \ref{sub:statistics}) and scaling laws of avalanches (Sec.\ \ref{sub:scaling})
with the focus on the influence of microscopic friction.
Then, we discuss the system size dependence of our results and explain the characteristic exponent for avalanche size distributions (Sec.\ \ref{sub:system_size}).
\subsection{Stress fluctuations}
\label{sub:stress}
We measure mechanical responses of the system to the applied strain $\gamma$ by the shear stress $\sigma$.
Neglecting kinetic contributions to the stress, we calculate the stress tensor according to
$\sigma_{\alpha\beta}=-L^{-2}\sum_{i<j}f_{ij\alpha}r_{ij\beta}$ ($\alpha,\beta=x,y$),
where $f_{ij\alpha}$ is the $\alpha$-component of the force between the particles ($i$ and $j$) in contact
and $r_{ij\beta}$ is the $\beta$-component of the relative position between them.
The shear stress is defined as the average of the off-diagonal elements,\ i.e.\ $\sigma\equiv(\sigma_{xy}+\sigma_{yx})/2$.
Note that the $\alpha$-component of the force consists of the normal and tangential forces,\ i.e.\ $f_n$ and $f_t$ (see Sec.\ \ref{sec:method}).
Accordingly, the shear stress can be decomposed into two parts as
\begin{equation}
\sigma = \sigma_n + \sigma_t~,
\label{eq:sigma_nt}
\end{equation}
where $\sigma_n$ and $\sigma_t$ are the normal and tangential parts of the shear stress, respectively
\footnote[1]{The stress tensor is decomposed as $\sigma_{\alpha\beta}=-L^{-2}\sum f_{ijn}n_{ij\alpha}r_{ij\beta}-L^{-2}\sum f_{ijt}t_{ij\alpha}r_{ij\beta}$,
where $n_{ij\alpha}$ ($t_{ij\alpha}$) is the $\alpha$-component of normal (tangential) unit vector.
The average of the off-diagonal elements of the first term (on the right-hand-side) is $\sigma_n$, whereas that of the second term is $\sigma_t$.}.
If the system is frictionless ($\mu=0$), we find $\sigma=\sigma_n$ and $\sigma_t=0$, whereas $\sigma_t\neq 0$ in frictional systems ($\mu>0$).
Because the tangential force does not exceed the threshold $\mu|f_n|$, $\sigma_t$ is smaller than $\sigma_n$ as long as $\mu<1$,
where we confirmed that the mean values in a steady state satisfy the following relation,
\begin{equation}
\langle\sigma_t\rangle\ll\langle\sigma_n\rangle\simeq\langle\sigma\rangle~.
\label{eq:mean_stress}
\end{equation}

\begin{figure}
\includegraphics[width=\columnwidth]{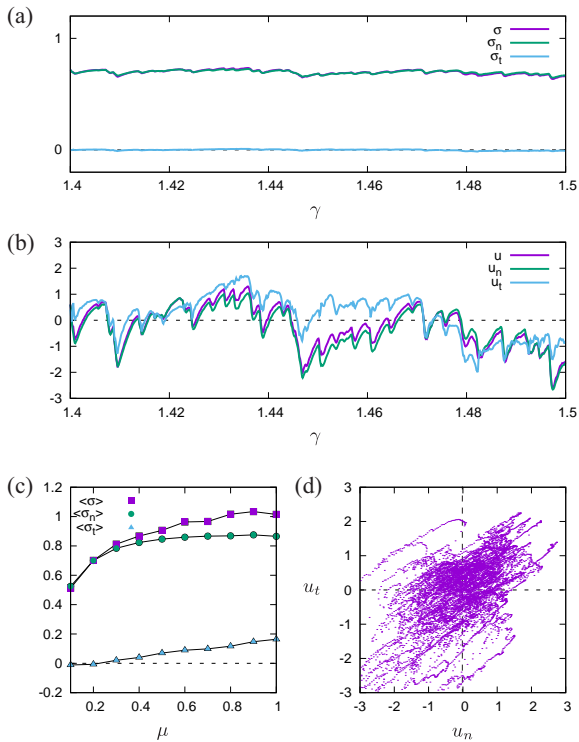}
\caption{
(a) Scaled fluctuations, $u$, $u_n$, and $u_t$, as functions of the strain $\gamma$.
(b) and (c): Scatter plots of (b) $u_n$ and $u$, and (c) $u_t$ and $u$, where each dot represents each value at $\gamma$.
Here, $N=8192$ and $\mu=0.2$ are used in the MD simulation.
\label{fig:stress-strain}}
\end{figure}

To study statistics of avalanches, we carefully look at fluctuations of the stress,\
i.e.\ $\delta\sigma\equiv\sigma-\langle\sigma\rangle$, $\delta\sigma_n\equiv\sigma_n-\langle\sigma_n\rangle$,
and $\delta\sigma_t\equiv\sigma_t-\langle\sigma_t\rangle$, in a steady state.
Figure \ref{fig:stress-strain}(a) displays the scaled fluctuations,
\begin{eqnarray}
u &=& \frac{\delta\sigma}{\sqrt{\langle\delta\sigma^2\rangle}}~,\\
u_n &=& \frac{\delta\sigma_n}{\sqrt{\langle\delta\sigma_n^2\rangle}}~,\\
u_t &=& \frac{\delta\sigma_t}{\sqrt{\langle\delta\sigma_t^2\rangle}}~,
\end{eqnarray}
as functions of the strain in the range $1.4\le\gamma\le 1.5$.
The shape of $u_n$ is quite similar with that of $u$, while the tangential part $u_t$ fluctuates almost independently of the others.
As can be seen in a scatter plot of $u$ and $u_n$ (Fig.\ \ref{fig:stress-strain}(b)), the normal part is strongly correlated with the total shear stress.
On the other hand, $u$ and $u_t$ are uncorrelated (Fig.\ \ref{fig:stress-strain}(c)) such that $\sigma_n$ and $\sigma_t$ are almost independent of each other.
%
\subsection{Stress drop events}
\label{sub:drop}
As shown in Fig.\ \ref{fig:stress-strain}(a),
the stress in a steady state fluctuates around its mean value such that numerous stress drop events, or \emph{avalanches}, can be observed.
To quantify such stress drop events, we calculate a stress drop amplitude as $\Delta\sigma\equiv\sigma(\gamma)-\sigma(\gamma+T)>0$,
where the shear stress starts decreasing at the strain $\gamma$ and stops decreasing at $\gamma+T$ \cite{aval_hatano}.
Here, $T$ is the so-called \emph{avalanche duration} \cite{aval_mean0,aval_mean5,aval_mean6}
and the \emph{avalanche size} is introduced as an extensive quantity, $S\equiv L^2\Delta\sigma$, as usual \cite{aval_inert0,aval_inert1,aval_inert2,aval_rate0}.

The avalanche duration $T$ is also defined such that the stress drop rate is negative,\
i.e.\ $d\sigma/d\gamma<0$, in the strain range between $\gamma$ and $\gamma+T$ \cite{aval_hatano}.
Therefore, we can extract the \emph{maximum stress drop rate} during an avalanche \cite{aval_mean5,aval_mean7} as
\begin{equation}
M \equiv -\left(\frac{d\sigma}{dt}\right)_\mathrm{max}~.
\label{eq:M}
\end{equation}
We also introduce the \emph{avalanche interval} $\Delta\gamma$ such that the stress drop rate is positive,\
i.e.\ $d\sigma/d\gamma>0$, in the strain interval between $\gamma$ and $\gamma+\Delta\gamma$.
%
\subsection{Statistics of avalanches}
\label{sub:statistics}
To clarify the effects of microscopic friction on statistics of avalanches,
we examine the MF predictions of avalanche size distributions (Sec.\ \ref{subsub:P(S)}),
distributions of the avalanche duration (Sec.\ \ref{subsub:P(T)}), and power spectra of stress drop rates (Sec.\ \ref{subsub:F(omega)}).
\subsubsection{Avalanche size distributions}
\label{subsub:P(S)}
First, we analyze statistics of avalanche sizes $S$.
Figure \ref{fig:PDF_FR-dependence}(a) displays probability distribution functions (PDFs) of avalanche sizes, or \emph{avalanche size distributions}, $P(S)$,
where the friction coefficient $\mu$ increases as indicated by the arrow (and listed in the legend).
If the avalanche size is not too small ($S>10^{-4}$), the avalanche size distributions exhibit the power-law decay,
\begin{equation}
P(S) \sim S^{-\tau}
\label{eq:P(S)}
\end{equation}
with the exponent $\tau$.
The power law (Eq.\ (\ref{eq:P(S)})) is followed by a ``bump" in large avalanche sizes ($S>1$)
due to the particle inertia as reported in Refs.\ \cite{aval_inert0,aval_inert1,aval_inert2}.
Therefore, the \emph{scaling regime} \cite{aval_mean6}, where the avalanche size distributions exhibit the power law, is in the range $10^{-4}<S<1$ in our MD simulations.

The exponent $\tau$ has been extensively studied by the theories \cite{aval_mean0,aval_mean1,aval_mean2,aval_mean3,aval_mean9},
experiments \cite{aval_mean5,aval_mean6,aval_mean7,aval_mean8,dahmen0},
and numerical simulations \cite{aval_foam0,aval_foam1,aval_quasi1,aval_finite0,aval_inert0,aval_inert1,aval_review0} (as reviewed in Sec. \ref{sec:intro}).
Here, we estimate $\tau$ by fitting the power law (Eq.\ (\ref{eq:P(S)})) to the data of $P(S)$ in the scaling regime, $10^{-4}<S<1$.
We find that $\tau\simeq 0.98$ (dotted line) for frictionless particles ($\mu=0$), while $\tau\simeq 1.33$ (solid line) for the frictional system with $\mu=0.1$.
The exponent for the frictionless system agrees with the previous result of soft athermal particles under shear \cite{aval_quasi1}.
On the other hand, that for the frictional particles is much closer to the MF prediction,\ i.e.\ $\tau=3/2$ \cite{aval_mean0,aval_mean1,aval_mean3,aval_mean6,aval_mean7,aval_mean8,dahmen0},
and results of EP models,\ i.e.\ $\tau\simeq 1.35$ \cite{aval_scaling0,aval_scaling1}.
Increasing the friction coefficient from $\mu=0$ to $0.1$,
we observe that the avalanche size distribution develops a ``plateau" in small avalanches, $S<10^{-4}$ (Fig.\ \ref{fig:PDF_FR-dependence}(a)).
Meanwhile, the exponent exhibits a transition from $\tau\simeq 0.98$ to $1.33$.
Therefore, the microscopic friction strongly affects the statistics of avalanche sizes and seems to play an important role in the MF approach to avalanches \cite{dahmen1}.
\subsubsection{Distributions of the avalanche duration}
\label{subsub:P(T)}
Next, we examine statistics of the avalanche duration $T$.
The MF theory predicts that PDFs of the avalanche duration show the power-law decay,
\begin{equation}
P(T) \sim T^{-\kappa}
\label{eq:P(T)}
\end{equation}
with the exponent $\kappa=2$ \cite{aval_mean0,aval_mean6}.
Figure \ref{fig:PDF_FR-dependence}(b) displays our numerical results of the PDFs,
where we increase the friction coefficient $\mu$ as indicated by the arrow (and listed in the legend).
In this figure, we can hardly see the power-law behavior of $P(T)$ in frictionless particles ($\mu=0$).
However, increasing $\mu$, we observe that the PDFs change their shapes and exhibit the power law (solid line) in the case of $\mu=0.1$.
We estimate the exponent as $\kappa=1.30$ by fitting the power law (Eq.\ (\ref{eq:P(T)})) to the data of $P(T)$ in the scaling regime, $10^{-6}<T<10^{-4}$.
Note that our exponent, $\kappa=1.30$, is smaller than the MF prediction, $\kappa=2$ \cite{aval_mean0,aval_mean6}.
In addition, the power law (Eq.\ (\ref{eq:P(T)})) is followed by a bump in the large avalanche duration ($T>10^{-4}$)
as in the case of avalanche size distributions (Fig.\ \ref{fig:PDF_FR-dependence}(a)).
\subsubsection{Power spectra of stress drop rates}
\label{subsub:F(omega)}
\begin{figure*}[t]
\includegraphics[width=\textwidth]{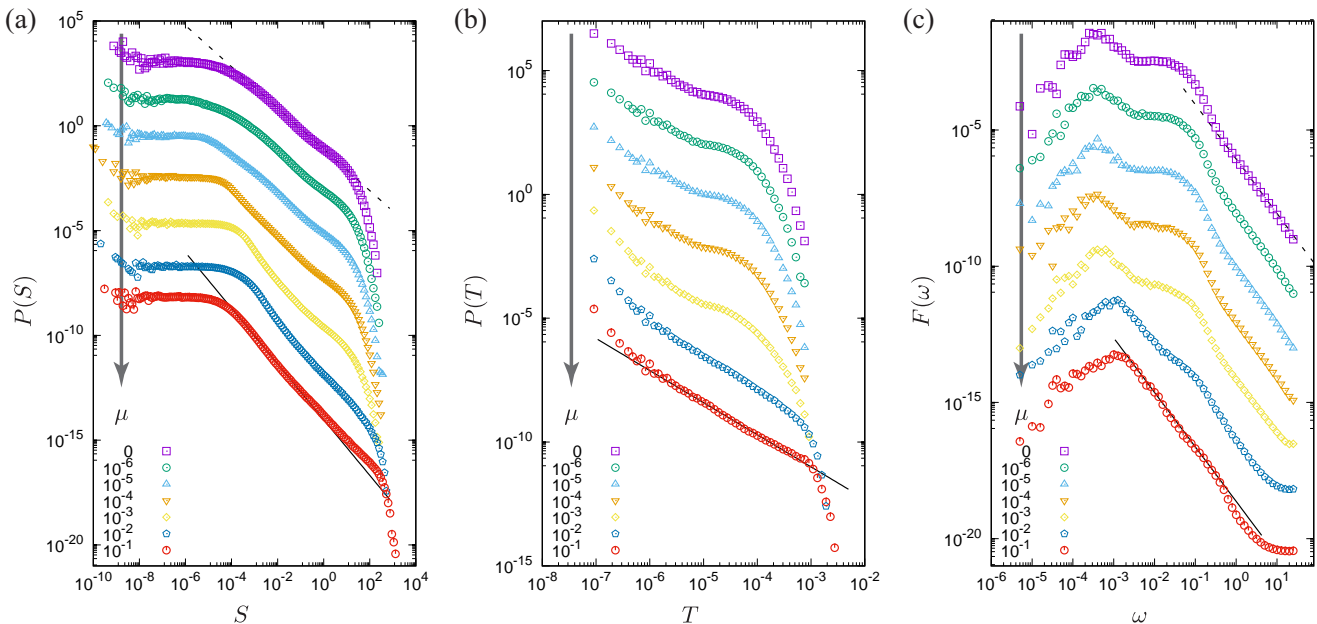}
\caption{
Double logarithmic plots of (a) avalanche size distributions $P(S)$, (b) distributions of the avalanche duration $P(T)$, and (c) power spectra of stress drop rates $F(\omega)$,
where we use $N=8192$ and increase the friction coefficient $\mu$ as indicated by the arrows and listed in the legends.
The dashed (solid) line in (a) indicates the power-law decay $P(S)\sim S^{-\tau}$ (Eq.\ (\ref{eq:P(S)})) with the exponent $\tau\simeq 0.98$ ($1.33$).
The solid line in (b) represents $P(T)\sim T^{-\kappa}$ (Eq.\ (\ref{eq:P(T)})) with $\kappa=1.30$.
Both the dashed and solid lines in (c) have the slope $-2$ (Eq.\ (\ref{eq:F(omega)})).
Note that the data for $\mu>0$ are arbitrarily shifted downward from the original positions to prevent overlap.
\label{fig:PDF_FR-dependence}}
\end{figure*}

We also examine the power spectrum of stress drop rate $F(\omega)\equiv|\hat{f}(\omega)|^2$,
where $\hat{f}(\omega)=\int_0^\infty(-d\sigma/dt)e^{-i\omega t}dt$ with the frequency $\omega$
is the Fourier transform of the stress drop rate in a steady state \cite{aval_spectrum}.
Figure \ref{fig:PDF_FR-dependence}(c) shows our numerical results of the power spectra,
where both the dashed and solid lines represent the MF prediction,\ i.e.\
\begin{equation}
F(\omega) \sim \omega^{-1/\sigma\nu z}
\label{eq:F(omega)}
\end{equation}
with the exponent $1/\sigma\nu z=2$ \cite{aval_mean0,aval_mean1,aval_mean6,aval_mean7}.
Increasing the friction coefficient $\mu$ (as indicated by the arrow and listed in the legend),
we observe that the lower bound of the power law (Eq.\ (\ref{eq:F(omega)})),\ i.e.\ the \emph{corner frequency} \cite{corner_f}, decreases.
For instance, the power-law decay begins around $\omega=10^{-1}$ ($10^{-3}$) if $\mu=0$ ($0.1$).
Therefore, the scaling regime for Eq.\ (\ref{eq:F(omega)}) shifts to lower frequencies in frictional particles.
%
\subsection{Scaling laws of avalanches}
\label{sub:scaling}
\begin{figure*}[t]
\includegraphics[width=\textwidth]{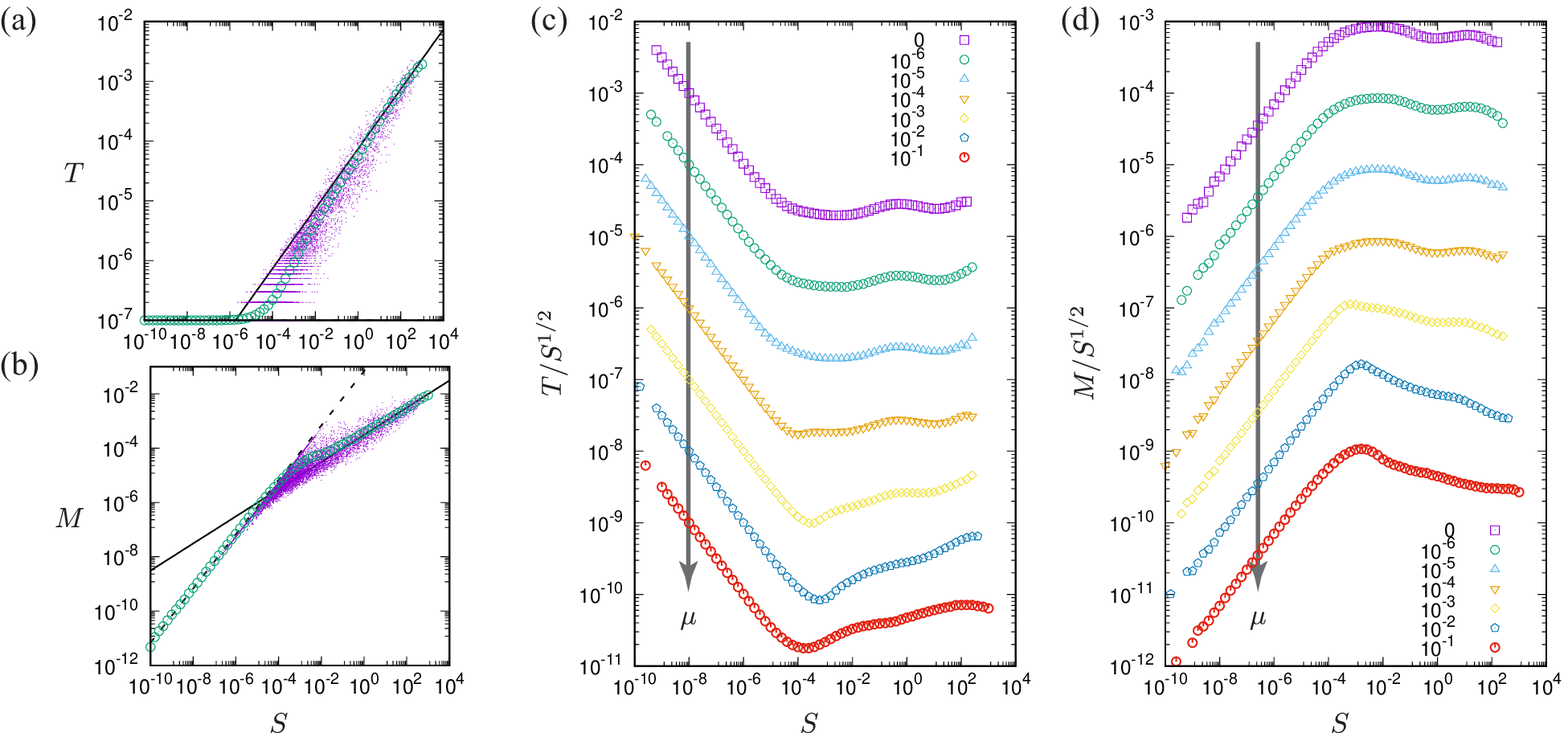}
\caption{
(a) and (b): Scatter plots of (a) the avalanche duration $T$ and (b) maximum stress drop rate $M$ as functions of the avalanche size $S$.
The data (each dot) are taken from $10^6$ stress drop events in a steady state, where we use $N=8192$ and $\mu=0.2$.
The symbols (circles) are the averages of (a) $T$ and (b) $M$ in each bin of $S$.
The solid lines in (a) and (b) represent the scaling laws,\ Eqs.\ (\ref{eq:T_S12}) and (\ref{eq:M_S12}), respectively.
The lower bound of avalanche duration is the strain increment,\ i.e.\ $T\geq\delta\gamma=10^{-7}$,
and the maximum stress drop rates for small avalanches are bounded by the linear relation, $M\propto S$ (dashed line in (b)).
(c) and (d): Double logarithmic plots of (c) $T/S^{1/2}$ and (d) $M/S^{1/2}$ as functions of $S$, where we use $N=8192$ and increase the friction coefficient $\mu$ as indicated by the arrows and listed in the legends.
Note that the data for $\mu>0$ are arbitrarily shifted downward from the original positions to prevent overlap.
\label{fig:TMS_FR-dependence}}
\end{figure*}

The MF theory also predicts that both the avalanche duration $T$ and maximum stress drop rate $M$ scale as the square root of the avalanche size $S$,\ i.e.\
\begin{eqnarray}
T &\sim& S^{1/2}~,\label{eq:T_S12}\\
M &\sim& S^{1/2}~.\label{eq:M_S12}
\end{eqnarray}
The MF predictions (Eqs.\ (\ref{eq:T_S12}) and (\ref{eq:M_S12})) were confirmed by experiments of bulk metallic glasses and granular materials \cite{aval_mean5}.
We also examine Eqs.\ (\ref{eq:T_S12}) and (\ref{eq:M_S12}) with the focus on the influence of microscopic friction.
In Figs.\ \ref{fig:TMS_FR-dependence}(a) and (b), we test the scaling laws of $T$ and $M$, respectively, for the case of frictional system with $\mu=0.2$.
Here, the data of $S$, $T$, and $M$ are taken from $10^6$ stress drop events in a steady state and the symbols (circles) are the averages of (a) $T$ and (b) $M$ in each bin of $S$.
The solid lines represent the scaling laws,\ Eqs.\ (\ref{eq:T_S12}) and (\ref{eq:M_S12}), which well describe our numerical results in the scaling regime, $10^{-2}<S<10^{2}$.
In Fig.\ \ref{fig:TMS_FR-dependence}(a), the lower bound of the avalanche duration is given by the strain increment,\ i.e.\ $T\geq\delta\gamma=10^{-7}$.
If the avalanche duration equals $\delta\gamma$,
the maximum stress drop rate is given by $M=\Delta\sigma/\delta t=S/L^2\delta t$ (see Eq.\ (\ref{eq:M})) so that $M$ is linear in the avalanche size $S$.
Therefore, the maximum stress drop rates for small avalanches are bounded by the linear scaling, $M\propto S$ (dashed line in Fig.\ \ref{fig:TMS_FR-dependence}(b)).

To examine the influence of microscopic friction on the scaling laws (Eqs.\ (\ref{eq:T_S12}) and (\ref{eq:M_S12})),
we plot the ratios, $T/S^{1/2}$ and $M/S^{1/2}$,\ i.e.\ \emph{shape indexes} \cite{aval_mean5}, in Figs.\ \ref{fig:TMS_FR-dependence}(c) and (d), respectively.
In these figures, we vary the friction coefficient from $\mu=0$ to $0.1$ as indicated by the arrow (and listed in the legend).
The ratios are almost flat in the scaling regime ($10^{-2}<S<10^{2}$) regardless of $\mu$ so that the scaling laws are quite insensitive to the microscopic friction.
However, $T/S^{1/2}$ ($M/S^{1/2}$) exhibits a ``dip" (``peak") around $S=10^{-4}$ when $\mu$ increases.
Therefore, the scaling law overestimates (underestimates) $T$ ($M$) in frictional systems outside of the scaling regime, $S<10^{-2}$.
%
\subsection{System size dependence and the exponent for avalanche size distributions}
\label{sub:system_size}
%
As shown in the previous sections,
the most striking difference between frictionless ($\mu=0$) and frictional ($\mu>0$) systems is the exponent $\tau$ for the avalanche size distributions $P(S)$.
To clarify the origin of this difference,
we study the system size dependence of avalanches in frictionless (Sec.\ \ref{subsub:frictionless}) and frictional particles (\ref{subsub:frictional}).
\subsubsection{Frictionless particles}
\label{subsub:frictionless}
\begin{figure}
\includegraphics[width=\columnwidth]{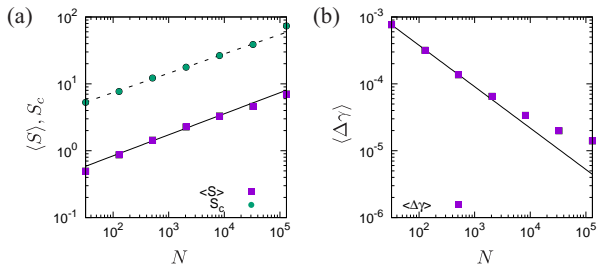}
\caption{
(a) The mean avalanche size $\langle S\rangle$ (squares) and cutoff size $S_c$ (circles),
and (b) the mean avalanche interval $\langle\Delta\gamma\rangle$ (squares) as functions of the number of frictionless particles $N$.
The lines indicate the power laws, $\langle S\rangle\sim N^\alpha$, $S_c\sim N^{d_f/d}$, and $\langle\Delta\gamma\rangle\sim N^{-\chi}$,
where we estimate the exponents as $\alpha\simeq 0.31$, $\chi\simeq 0.62$, and $d_f/d\simeq 0.29$ (i.e.\ $d_f\simeq 0.58$).
\label{fig:ave_frictionless}}
\end{figure}

To explain the exponent $\tau$ for the frictionless system ($\mu=0$),
we demonstrate the \emph{finite size scaling} of the mean avalanche size $\langle S\rangle$ and interval $\langle\Delta\gamma\rangle$,
which has been well established by EP models \cite{aval_scaling0,aval_scaling1,aval_scaling2}.
We calculate the mean values, $\langle S\rangle$ and $\langle\Delta\gamma\rangle$, by taking the averages of avalanche size and interval over $10^6$ stress drop events in a steady state.
Figure \ref{fig:ave_frictionless} displays (a) $\langle S\rangle$ and (b) $\langle\Delta\gamma\rangle$ in the frictionless system as functions of the system size (the number of particles) $N$,
where we confirm the power laws, $\langle S\rangle\sim N^\alpha$ and $\langle\Delta\gamma\rangle \sim N^{-\chi}$ (solid lines).
The power-law exponents are given by $\alpha\simeq 0.31$ and $\chi\simeq 0.62$, which reasonably agree with the \emph{scaling relation} in Refs.\ \cite{aval_scaling2},
\begin{equation}
\alpha + \chi = 1~.
\label{eq:scaling_relation1}
\end{equation}

We also calculate the \emph{cutoff size} as \cite{aval_scaling2}
\begin{equation}
S_c \equiv \frac{\langle S^2\rangle}{\langle S\rangle}~.
\label{eq:S_c}
\end{equation}
As shown in Fig.\ \ref{fig:ave_frictionless}(a) (dotted line),
the cutoff size exhibits the power law $S_c \sim N^{d_f/d}$ with the spatial dimension $d=2$ and fractal dimension $d_f\simeq 0.58$.
By using another scaling relation \cite{aval_scaling2},
\begin{equation}
\tau = 2-\alpha\frac{d}{d_f}~,
\label{eq:scaling_relation2}
\end{equation}
we can estimate the exponent $\tau$ from $\alpha$ and $d_f$.
We find $\tau\simeq 0.93$ from Eq.\ (\ref{eq:scaling_relation2})
which reasonably agrees with $\tau\simeq 0.98$ obtained by fitting Eq.\ (\ref{eq:P(S)}) to the data of $P(S)$ (dotted line in Fig.\ \ref{fig:PDF_FR-dependence}(a)).
Therefore, the exponent $\tau$ for avalanche size distributions in the frictionless system
can be explained by the finite size scaling of $\langle S\rangle$, $\langle\Delta\gamma\rangle$, and $S_c$ \cite{aval_scaling0,aval_scaling1,aval_scaling2}.
%
%
\subsubsection{Frictional particles}
\label{subsub:frictional}
The system size dependence of the mean avalanche size $\langle S\rangle$ and interval $\langle\Delta\gamma\rangle$ in frictional systems is entirely different from that in the frictionless system.
In Appendix \ref{app:scaling_friction}, we examine the dependence of the mean values, $\langle S\rangle$ and $\langle \Delta\gamma\rangle$, on the system size $N$ and friction coefficient $\mu$.
We find that their dependence on $N$ is sensitive to $\mu$ so that the scaling relations established for the case of $\mu=0$
(Eqs.\ (\ref{eq:scaling_relation1}) and (\ref{eq:scaling_relation2})) are not applicable to frictional particles, $\mu>0$ (see Appendix \ref{app:scaling_friction}).

\begin{figure}
\includegraphics[width=\columnwidth]{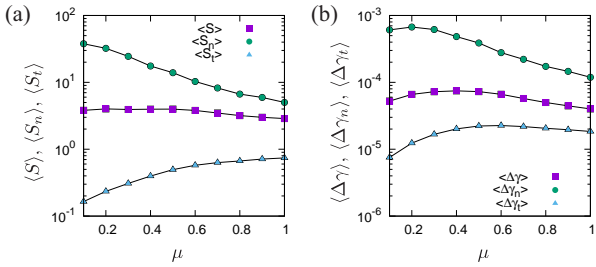}
\caption{
Semi-logarithmic plots of (a) the mean avalanche sizes, $\langle S\rangle$, $\langle S_n\rangle$, and $\langle S_t\rangle$,
and (b) mean avalanche intervals, $\langle\Delta\gamma\rangle$, $\langle\Delta\gamma_n\rangle$, and $\langle\Delta\gamma_t\rangle$,
as functions of the friction coefficient $\mu$.
\label{fig:mean}}
\end{figure}

To reveal how the microscopic friction alters the exponent $\tau$ for avalanche size distributions $P(S)$, we analyze normal and tangential contributions to the avalanche size $S$.
As shown in Sec.\ \ref{sub:stress},
the shear stress $\sigma$ is divided into the normal and tangential parts as $\sigma_n$ and $\sigma_t$, respectively (Eq.\ (\ref{eq:sigma_nt})).
Accordingly, we distinguish the avalanche size as $S_n\equiv L^2\Delta\sigma_n$ and $S_t\equiv L^2\Delta\sigma_t$, and avalanche interval as $\Delta\gamma_n$ and $\Delta\gamma_t$.
Figure \ref{fig:mean}(a) displays the mean avalanche sizes, $\langle S\rangle$, $\langle S_n\rangle$, and $\langle S_t\rangle$, as functions of the friction coefficient $\mu$.
As can be seen, the mean avalanche size of normal part is one to two decades larger than that of tangential part, where a relation,
\begin{equation}
\log_{10}\langle S_t\rangle < \log_{10}\langle S\rangle < \log_{10}\langle S_n\rangle~,
\label{eq:average_d}
\end{equation}
holds regardless of $\mu$ (we use $\log_{10}$ to notice the difference in logscale).
In frictional systems, microscopic ``slip" between the particles in contact,\
i.e.\ the switch from the tangential force $|f_t|$ to Coulomb friction $\mu|f_n|$, is responsible for the avalanche of tangential part $S_t$.
If $\mu$ is small, the frictional particles can easily ``slip" so that the energy to be released at a tangential stress drop is small.
Thus, the mean avalanche size $\langle S_t\rangle$ decreases with the decrease of $\mu$ as shown in Fig.\ \ref{fig:mean}(a).
On the other hand, the mean size of normal part $\langle S_n\rangle$ increases with the decrease of $\mu$
such that the mean avalanche size $\langle S\rangle$ is quite insensitive to the strength of microscopic friction.
In addition, the mean avalanche intervals satisfy
\begin{equation}
\log_{10}\langle\Delta\gamma_t\rangle < \log_{10}\langle\Delta\gamma\rangle < \log_{10}\langle\Delta\gamma_n\rangle
\label{eq:average_f}
\end{equation}
regardless of $\mu$ (Fig.\ \ref{fig:mean}(b)).
Therefore, microscopic slips which trigger small avalanches of tangential part $S_t$ are more frequent than normal stress drop events.
This means that numerous small avalanches happen in frictional systems, which significantly increases the population of small $S$ in the avalanche size distribution $P(S)$.
Accordingly, the exponent $\tau$ becomes larger (and closer to the MF prediction) than that in the frictionless case.

\begin{figure}
\includegraphics[width=\columnwidth]{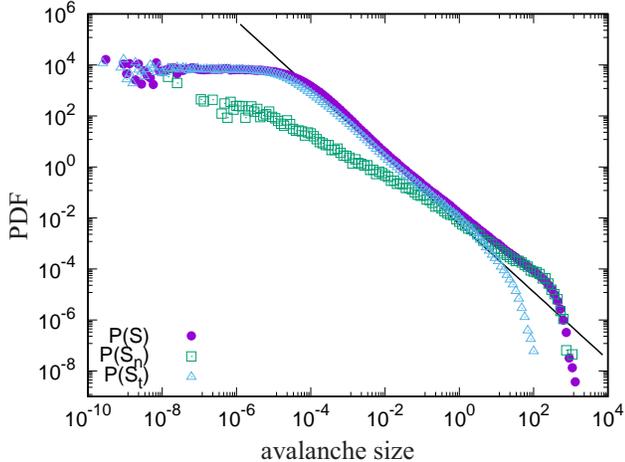}
\caption{
Double logarithmic plots of the avalanche size distribution, $P(S)$, and PDFs of normal and tangential avalanches, $P(S_n)$ and $P(S_t)$,
where the PDFs ($P(S_n)$ and $P(S_t)$) are arbitrarily shifted downward to be compared with $P(S)$.
The solid line indicates the power-law decay (Eq.\ (\ref{eq:P(S)})) with the exponent $\tau=1.33$ as in Fig.\ \ref{fig:PDF_FR-dependence}(a).
The system size and friction coefficient are given by $N=8192$ and $\mu=0.1$, respectively.
\label{fig:P(S_n_t)}}
\end{figure}

To confirm our interpretation on the exponent $\tau$ for frictional systems,
we compare the PDFs of normal and tangential parts,\ i.e.\ $P(S_n)$ and $P(S_t)$, with the avalanche size distribution, $P(S)$.
Figure \ref{fig:P(S_n_t)} displays the PDFs, $P(S)$, $P(S_n)$, and $P(S_t)$, where we use $N=8192$ frictional particles with $\mu=0.1$
(note that a similar analysis has recently been proposed in Ref.\ \cite{aval_oyama2020}).
In this figure, both the plateau of $P(S)$ in small avalanches ($S<10^{-4}$) and power-law decay of $P(S)$ in the scaling regime ($10^{-4}<S<1$)
are well represented by the PDF of tangential part $P(S_t)$,
where the solid line is the power-law decay (Eq.\ (\ref{eq:P(S)})) with the exponent $\tau=1.33$ as in Fig.\ \ref{fig:PDF_FR-dependence}(a).
On the other hand, both the bump of $P(S)$ in large avalanches ($S>1$) and tail of $P(S)$ correspond to those of the PDF of normal part $P(S_n)$.
Because the normal and tangential parts of the stress are uncorrelated (Sec.\ \ref{sub:stress}),
we crudely estimate $P(S)\sim P(S_t)+P(S_n)$,
where the power-law exponent is solely determined by the PDF of tangential part,\ i.e.\ $P(S_t)\sim S_t^{-\tau}$.
%
%
\section{Discussion}
\label{sec:discuss}
In this study, we have numerically investigated stress drop events, or avalanches, in two-dimensional frictional particles under shear.
Our focus was the effect of microscopic friction on statistics (Sec.\ \ref{sub:statistics}) and scaling laws of avalanches (Sec.\ \ref{sub:scaling}).
Based on the MF predictions, we examined the avalanche size distributions $P(S)$, PDFs of the avalanche duration $P(T)$, and power spectra of stress drop rates $F(\omega)$.
We found that the avalanche size distributions exhibit the power law, $P(S)\sim S^{-\tau}$ (Eq.\ (\ref{eq:P(S)})), in the scaling regime $10^{-4}<S<1$,
where the power-law exponent for frictional particles, $\tau=1.33$, is larger and closer to the MF prediction ($\tau=3/2$)
and results of the EP models ($\tau=1.35$) than that for frictionless particles, $\tau=0.98$.
If the friction coefficient is finite $\mu>0$,
the PDFs of the avalanche duration also show the power-law decay $P(T)\sim T^{-\kappa}$ (Eq.\ (\ref{eq:P(T)})) in the scaling regime $10^{-6}<T<10^{-4}$
though the exponent $\kappa=1.30$ is smaller than the MF value $\kappa=2$.
On the other hand, we cannot observe the power-law behavior of $P(T)$ if the system is frictionless $\mu=0$.
The MF theory well explains our numerical results of the power spectra,\ i.e.\ $F(\omega)\sim\omega^{-2}$, regardless of $\mu$.
However, we found that the lower bound of the power law (i.e.\ the corner frequency) decreases with the increase of $\mu$
such that the scaling regime (for the spectra) shifts to lower frequencies in frictional systems.
We also examined the scaling laws of the avalanche duration and maximum stress drop rate,\ i.e.\ $T\sim S^{1/2}$ and $M\sim S^{1/2}$ (Eqs.\ (\ref{eq:T_S12}) and (\ref{eq:M_S12})),
and found that both are correct in the scaling regime $10^{-2}<S<10^2$ regardless of $\mu$.
The influence of microscopic friction emerges outside of the scaling regime (around $S=10^{-4}$);
the small dip and peak appear in the shape indexes, $T/S^{1/2}$ and $M/S^{1/2}$, respectively.
The most striking difference between the frictionless and frictional systems is the different exponents for the avalanche size distributions, $\tau$.
To clarify the origin of the difference, we demonstrated the finite size scaling (Sec.\ \ref{sub:system_size}).
If the system is frictionless $\mu=0$, the exponent $\tau$ is well explained by the scaling relations between the spatial dimension, fractal dimension,
and power-law exponents for the mean avalanche size $\langle S\rangle$ and interval $\langle\Delta\gamma\rangle$ (Eqs.\ (\ref{eq:scaling_relation1}) and (\ref{eq:scaling_relation2})).
However, the scaling relations do not hold in frictional systems $\mu>0$ and cannot explain the exponent $\tau$ which is closer to the MF prediction.
To understand the exponent for frictional systems, we analyzed the avalanches of normal and tangential parts of the shear stress.
We found that the tangential part of the stress contributes to small sized avalanches.
In addition, the avalanches of tangential part are more frequent than those of normal part.
Because the normal and tangential parts are uncorrelated (Sec.\ \ref{sub:stress}),
we concluded that numerous small avalanches of the tangential stress increase the slope of the avalanche size distribution,
which eventually increases the power-law exponent $\tau$ for frictional particles.
\subsection{The role of microscopic friction in the MF approach to avalanches}
From our numerical results, we find that the statistics of avalanches (especially the exponent $\tau$) in frictional systems are close to the MF prediction and results of the EP models.
Both the MF theory \cite{aval_mean0,aval_mean1,aval_mean2,aval_mean3} and EP models \cite{aval_review0,aval_inert2,aval_rate0} are based on non-local constitutive equations of stress,
where the elastic propagator of local deformations is assumed to be long-ranged (though anisotropic quadrupolar symmetry is introduced to that in the EP models).
It is known that, if the microscopic friction is present, the system exhibits a sudden increase of stress
such as \emph{shear jamming} \cite{s-jamming0} and \emph{discontinuous shear thickening} \cite{DST0,DST1,DST2,DST3,DST4,DST5,DST6,DST7,DST8},
where force chains are percolated through the system under shear.
The percolated force chains are reminiscent of long-range spatial correlations and analogous to the propagator in the MF theory and EP models.
Therefore, we suppose that the microscopic friction is a key ingredient of the MF approach and plays a crucial role in avalanches observed in real materials.
%
\subsection{Future works}
In our MD simulations, we fixed the area fraction of the particles to $\phi=0.9$ which is far above the \emph{jamming transition density} $\phi_J\simeq 0.8433$ \cite{gn3,gn4,rheol0}.
However, mechanical properties of soft athermal particles drastically change if the system approaches the onset of unjamming,\
i.e.\ $\phi\rightarrow\phi_J$ \cite{tanguy0,tanguy1,tanguy2,tanguy3,rs0,rs1}.
Thus, it is an important next step to study how the avalanche statistics and scaling laws are affected by the proximity to jamming.
We also fixed the strain increment to $\delta\gamma=10^{-7}$, where the shear rate is given by $\dot{\gamma}\equiv\delta\gamma/\delta t=10^{-6}t_0^{-1}$ in our unit.
However, it is known that the shape of avalanche size distribution \cite{aval_rate1}, as well as the exponent $\tau$ \cite{aval_rate0}, is sensitive to $\dot{\gamma}$.
Therefore, it is also important to investigate the effects of driving rate \cite{aval_mean4} or time resolution \cite{aval_mean10} on our numerical results.
We employed the spring-dashpot model for the interaction between the particles in contact (Sec.\ \ref{sec:method}) as a canonical model of granular materials \cite{dem}.
However, interaction forces drastically change the flow behavior of soft athermal particles,\
e.g.\ the force law of viscous damping controls the shear thinning/thickening \cite{rheol7} and the flow curves are nonmonotonic if one introduces cohesive forces \cite{AST0,AST1}.
Therefore, it is interesting to examine how the interaction forces alter the avalanche statistics and scaling laws.
In addition, we had not analyzed spatial structures of the system such as \emph{non-affine displacements} of the particles
\cite{pdf0,pdf1,pdf2,corl0,corl1,corl2,corl3,corl7,corl8,spectrum,Combe,saitoh11,saitoh12}
and spatial correlations of the stress \cite{aval_stress_corr0,aval_stress_corr1,aval_stress_corr2,aval_stress_corr3}.
Thus, detailed studies of the link between statistics and spatial structures are left for future.
Similarly, anisotropy induced by shear \cite{aniso0,aniso1,aniso2,aniso3,aniso4,aniso5,saitoh13},\ e.g.\ shear-bands, should be examined in future.
Moreover, further studies in three dimensions are crucial to practical applications of our results \cite{saitoh16}.
%
\begin{acknowledgments}
I thank Norihiro Oyama for fruitful discussions and helpful comments on the manuscript.
This work was financially supported by JSPS KAKENHI Grant Numbers 18K13464 and 20H01868.
\end{acknowledgments}
\appendix*
\section{Finite size scaling in frictional systems}
\label{app:scaling_friction}
In this appendix, we explain how the microscopic friction alters the finite size scaling established for the frictionless system and discuss the \emph{frictionless limit}.

\begin{figure}
\includegraphics[width=\columnwidth]{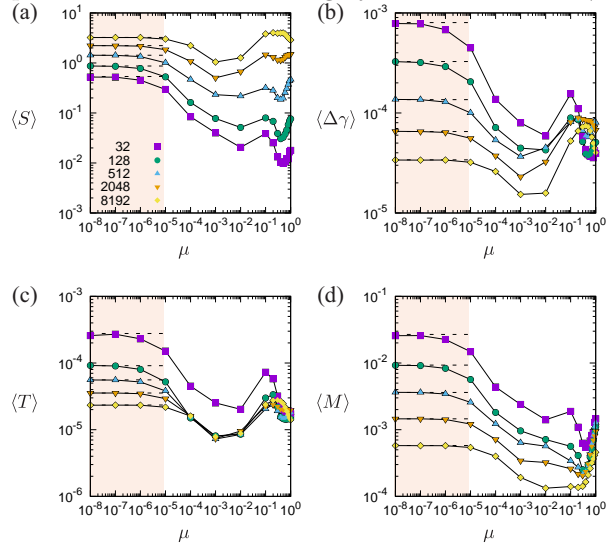}
\caption{
Double logarithmic plots of (a) the mean avalanche size $\langle S\rangle$, (b) mean avalanche interval $\langle\Delta\gamma\rangle$,
(c) mean avalanche duration $\langle T\rangle$, and (d) mean maximum stress drop rate $\langle M\rangle$ as functions of the friction coefficient $\mu$.
The symbols represent the system size $N$ as listed in the legend of (a).
The horizontal dashed lines indicate the values in the frictionless system, $\mu=0$,
while the shaded regions indicate the \emph{frictionless limit}, $\mu<\mu_c\sim 10^{-5}$.
\label{fig:ave_frictional}}
\end{figure}

Figure \ref{fig:ave_frictional} displays (a) the mean avalanche size $\langle S\rangle$ and (b) mean avalanche interval $\langle\Delta\gamma\rangle$
as functions of the friction coefficient $\mu>0$, where we change the system size $N$ as listed in the legend of (a).
As can be seen, the system size dependence of $\langle S\rangle$ and $\langle\Delta\gamma\rangle$ is sensitive to $\mu$
so that the scaling relations established for the case of $\mu=0$ (Eqs.\ (\ref{eq:scaling_relation1}) and (\ref{eq:scaling_relation2})) are not applicable to $\mu>0$.
In this figure, we also show (c) the mean avalanche duration $\langle T\rangle$ and (d) mean maximum stress drop rate $\langle M\rangle$ as functions of $\mu$.
All the mean values significantly deviate from the frictionless cases (horizontal dashed lines) if the friction coefficient exceeds the critical value $\mu_c\sim 10^{-5}$.
Here, the critical value can be estimated as follows:
In each time step, the strain increment $\delta\gamma$ is applied to the system,
where the tangential displacement between the particles in contact is roughly estimated as $d_0\delta\gamma$ with the mean particle diameter $d_0$.
The particles ``slip" if the tangential force $f_t\sim k_t d_0\delta\gamma$ exceeds the threshold $\mu|f_n|\sim \mu k_n\xi$,\ i.e.\ if $k_t d_0\delta\gamma>\mu k_n\xi$.
Since $k_n=k_t$ and $\delta\gamma=10^{-7}$ in our MD simulations, this condition leads to $\mu<\mu_c\equiv d_0\delta\gamma/\xi\sim 10^{-5}$,
where the overlap between the particles is roughly $\xi\sim 10^{-2}d_0$ \cite{saitoh15}.
Because the critical value $\mu_c$ is infinitesimal and particles slip every time step, the system is almost frictionless in the limit, $\mu<\mu_c$ (shaded regions).
%
%
\bibliography{avalanche}
\end{document}